\documentclass[]{piparticle-final}
\usepackage{graphicx}
\usepackage{amsmath}
\usepackage{enumerate}

\usepackage{cite} 
\usepackage{epstopdf} 

\begin{document}

\volume{7}               
\articlenumber{070001}   
\journalyear{2015}       
\editor{C. S. O'Hern}   
\reviewers{M. Pica Ciamarra, Nanyang Technological University, Singapore.}  
\received{19 January 2015}     
\accepted{25 February 2015}   
\runningauthor{D. Slobinsky \itshape{et al.}}  
\doi{070001}         

\title{Wang--Landau algorithm for entropic sampling of arch-based microstates in the volume ensemble of static granular packings}

\author{D. Slobinsky,\cite{inst1,inst2}\thanks{E-mail: dslobinsky@frlp.utn.edu.ar} \hspace{0.5em} 
        Luis A. Pugnaloni\cite{inst1,inst2}\thanks{E-mail: luis.pugnaloni@frlp.utn.edu.ar}}

\pipabstract{
We implement the Wang--Landau algorithm to sample with equal probabilities the static configurations of a model granular system. The ``non-interacting rigid arch model'' used is based on the description of static configurations by means of splitting the assembly of grains into sets of stable arches. This technique allows us to build the entropy as a function of the volume of the packing for large systems. We make a special note of the details that have to be considered when defining the microstates and proposing the moves for the correct sampling in these unusual models. We compare our results with previous exact calculations of the model made at moderate system sizes. The technique opens a new opportunity to calculate the entropy of more complex granular models. 
}

\maketitle

\blfootnote{
\begin{theaffiliation}{99}
   \institution{inst1} Departamento de Ingenier\'ia Mec\'anica, Facultad Regional La Plata, Universidad Tecnol\'ogica Nacional, Av. 60 Esq. 124, 1900 La Plata, Argentina.
   \institution{inst2} Consejo Nacional de Investigaciones Cient\'ificas y T\'ecnicas (CONICET), Argentina.
\end{theaffiliation}
}

\section{Introduction}

In the study of static packings there exists still a lack of predictive capabilities of the available theories. Assemblies of objects that pack (such as grains) can generally sample such packed configurations only by the external excitation of the system. These packings can be built by repeating a given packing protocol (e.g., homogeneous compression or deposition under an external field against a confining boundary) on an initial random configuration. Also, a Markovian or non-Markovian series can be constructed by exciting the system from the previous packing configuration. To what extent the series of packings obtained (using either type of protocol) can be modeled without information on the dynamics that drives the system to the packed configuration is still uncertain. The main reason for this is that the few statistical approaches that attempt to do this are strongly hinder by the poor current ability to generate such packed structures without using a dynamics to build the packings.

One might expect that the packing fraction and its fluctuations, among other properties, could be obtained from basic statistics without resourcing to a full molecular dynamic type of simulations (also known as ``discrete element method'', DEM). Although these types of simulations are powerful enough to predict the behaviour of most systems that pack, it is desirable to find a description that could neglect the detailed dynamics between consecutive packed configurations.   

The use of the tools provided by the ensemble theory of statistical mechanics in problems of granular matter is still limited. Although many studies perform statistical analysis of configurations of a granular sample obtained during careful preparations in the laboratory or in molecular dynamic-type simulations, very rarely sampling in a particular statistical ensemble is carried out either via an analytic or a computational calculation of a model system. This has prevented a direct assessment as to whether ensemble theory is appropriate to describe the behaviour of these peculiar systems.

One pioneering contribution to the topic is the idea that granular systems at mechanical equilibrium could be treated as ensemble members, putting forward the conjecture that the mean values of measurable quantities could be calculated using statistical mechanics for these ensembles \cite{Edwards19891080}. In this scheme, each of the thermodynamic variables finds a counterpart, the {\it volume} taking the place of the energy, and the {\it compactivity} that of the temperature, amongst other transformations. However beautiful this description may seem, the computational challenges to generate ensemble samples in this context are extraordinaries \cite{mcnamara2009measurement,PhysRevLett.101.128001,PhysRevLett.112.098002,PhysRevE.68.011306}. 

One complication that prevents to a large extent the use of the machinery of statistical mechanics in this case is the fact that configurations, unlike in traditional liquid theories, have to be checked for the constraint of mechanical equilibrium. In a previous work \cite{archensemble}, we have made a proposal on how to deal with this, at least in a first approximation, by describing the excitations of static granular systems under gravity in terms of its arches. Since arches are sets of grains that stabilize each other, these are the basic units of mechanically stable structures in the packing. Any static configuration can be described in terms of the arches formed by its grains, their arch shape, position, orientations, etc. We have considered, as an example, a model of a two-dimensional (2D) granular system composed of disks where arches are assumed to take a single possible structure and the arch--arch interactions (due to the interlocking of arches) is neglected. We calculated the exact entropy of
this model (the non-interacting rigid arch model, NIRA) by constructing all possible configurations for moderate system sizes. Of course, generating each state is a cumbersome task if the system size is increased or if the number of degrees of freedoms (DoF) is increased by using more realistic models. Therefore, an alternative approach based on sampling the phase space with the desired probability is necessary.

In the present work, we will calculate the entropy of the NIRA model in the microcanonical ensemble \cite{PhysRevLett.71.211} using entropic sampling through the Wang--Landau (WL) algorithm \cite{PhysRevLett.86.2050,PhysRevE.64.056101,PhysRevE.72.025701,PhysRevE.70.046701}. This approach allows us to obtain the entire entropy function for all possible volumes of the system in one single simulation for larger systems and potentially for more complex models. All derived properties, such as compactivity and volume fluctuations, can then be calculated through numerical differentiation. We pay particular attention to the different descriptions that can be realized for the NIRA model. Some of these representations do not provide a direct way of sampling the configuration space uniformly.

This work is organized as follows: in section \ref{wl}, we will review the WL algorithm. In section \ref{ras}, we will review the NIRA model and discuss different ways of representing it, along with the issues related to uniform sampling of the configurations. We then present a representation that allows very fast calculations of the entropy and we compare the results with the exact counting of all configurations for systems of moderate size. Finally, we discuss future directions to refine the arch-based ensemble volume function towards capturing detailed features of more realistic systems.

\section{Wang-Landau algorithm}\label{wl}
The WL method has revolutionized computational statistical mechanics \cite{PhysRevLett.86.2050,PhysRevE.64.056101,PhysRevE.72.025701,PhysRevE.70.046701}.  WL is a pure statistical method that can retrieve the density of states (DoS) (hence the entropy) over a bounded region of the energy spectrum from the sole knowledge of the energy function. The spectacular computational performance achieved by this method stems from the fact that it presents no limitations for the system to tunnel between potential barriers, in stark contrast with classical Monte Carlo methods that underperform when they encounter deep valleys in the energy landscape.

WL finds the entropy of the system by means of a Markov chain in the energy landscape which is conveniently biased towards the less probable energies in a strongly history dependent manner. This is achieved by using the multicanonical approach \cite{PhysRevLett.68.9} in which each possible configuration is sampled with a probability given by the inverse of the density of states for its given energy.
Specifically, WL aims to obtain a flat histogram of visited energies $E$ by forcing the system to go through all configurations with a probability which is inverse to the previous occurrence of that energy in the Markov chain. The method is ergodic and asymptotically fulfils the detailed balance condition \cite{NewEntry12}. There exists extensive literature on the WL algorithm but here, we only summarize its most relevant steps. In the following sections, we will refer to a \emph{configuration} of the system as a fixed set of values of all its DoF. 

In WL, one defines two histograms that are continuously updated as the Markov chain proceeds. These histograms are the entropy $S(E)$, which is the output of the algorithm, and a control histogram $H(E)$. After initializing $H(E)=0$, $S(E)=1$ and a starting configuration with energy $E_0$, the rules to update these histograms are:

\begin{enumerate}[I]
    \item Propose a new configuration and calculate its energy $E_1$. The new configuration is generally derived from the previous configuration by a change in the value of one of its DoF.
    \item Accept the new configuration according to a probability given by: $\min\left[1,\exp(S(E_0)-S(E_1))\right]$. 
    \item Update the two histograms in the correct energy bin $E$ ($E_1$ if the new configuration is accepted and $E_0$ if otherwise), accordingly: $H(E)=H(E)+1$ for the control histogram, and $S(E)=S(E)+f$ for the entropy. Here, $f$ is a correction that controls the precision of the algorithm, which will be decreased (see next step), usually starting at $f=1$. 
    \item If the control histogram $H(E)$ is flat enough according to some arbitrary criterion, decrease $f$ (for instance by making $f=f/2$) and reset all entries of $H(E)$ to zero.
    \item If $f>\epsilon$ (with $\epsilon$ a prescribed tolerance), return to step I, otherwise stop.
\end{enumerate}

After each reduction of the correction term $f$, the entropy histogram is built with a finer grained precision. However, to speed up the initial estimates of $S(E)$, $f$ is set to a high initial value. Different approaches are followed to accelerate the final stages of refinement by decreasing $f$ with alternative criteria \cite{PhysRevE.75.046701}.

In Monte Carlo approaches, the detailed balance condition ensures that the Markov chain has a limiting distribution \cite{newman1999monte}. Detailed balance can be stated as follows:
\begin{equation}
    p_\mu P(\mu\rightarrow\nu) = p_\nu P(\nu\rightarrow\mu) \label{detailedbalance}
\end{equation}
where $p_\mu$ is the probability distribution of configuration $\mu$ and $P(\mu\rightarrow\nu)$ the transition probability from configuration $\mu$ to configuration $\nu$ which can be written as:
\begin{equation}
    P(\mu\rightarrow\nu) = S_P(\mu\rightarrow\nu)A_P(\mu\rightarrow\nu) \label{selection}
\end{equation}
with $S_P(\mu\rightarrow\nu)$ the selection probability, which is the probability that the algorithm generates a trial configuration $\nu$ starting from configuration $\mu$; and $A_P(\mu\rightarrow\nu)$ the acceptance probability, i.e., the probability that the algorithm will accept the trial configuration $\nu$. Hence, since the target distribution in the entropic sampling is the inverse of the DoS, i.e., $p_\mu \propto g(E_\mu)^{-1} = \exp[-S(E_\mu)]$, Eqs. (\ref{detailedbalance}) and (\ref{selection}) implies

\begin{equation}
    \frac{A_P(\mu\rightarrow\nu)}{A_P(\nu\rightarrow\mu)} = \exp[S(E_\mu)-S(E_\nu)] \frac{S_P(\nu\rightarrow\mu)}{S_P(\mu\rightarrow\nu)} \label{aceptance}
\end{equation}

In WL, detailed balance is not fulfilled in general. During the construction of entropy, the acceptance probability given in step II above (i.e., $A_P(\mu\rightarrow\nu)=\min\left[1,\exp(S(E_\mu)-S(E_\nu))\right]$) evolves and it is only when the entropy gets sufficiently refined that detailed balance is met. 

Notice that the form chosen for $A_P$ requires that the selection probability be symmetric (i.e., $S_P(\mu\rightarrow\nu)=S_P(\nu\rightarrow\mu)$). Although this condition is simple to comply with in models like Ising, for arch-based descriptions of static granular packs this is non-trivial. Therefore, one must be careful to represent a system in such a way that trial moves between different configurations have the same backward and forward selection probability. 

After reviewing the main characteristics of the arch-based ensemble in the next section, we will show that some natural representations of the microstates lead to non-symmetric selection probability schemes. We present, however, a way of representing the configurations that does allow for the direct use of WL. In all the WL simulations, we have used $300$ bins for the histograms. The tolerance for the $f$ correction was set to $\epsilon = 2^{-15}$. The histogram $H(E)$ is considered flat (see step IV) whenever $(H_{max}-H_{min})/H_{max}<0.2$, with $H_{max}$ and $H_{min}$ the maximum and minimum height of the histogram.

\section{Arch-based microstates}\label{ras}
In Ref. \cite{archensemble}, we have introduced a way of describing the microstate of a static granular system under gravity by considering the arches that the grains form instead of the more traditional approach of using the particle positions. Arches are defined as sets of mutually stable grains. All other particles being fixed, the removal of any of the grains in the arch would induce the destabilization of the rest of the set. Any assembly of grains, static under gravity, can be split into a number of arches which are mutually exclusive \cite{pugnaloni,arevalo}.

The major difficulty in sampling static granular configurations is the fact that these are sparse (with zero measure) in the overwhelming number of possible particle positions. Moreover, there are not recipes to generate a static configuration from another static configuration by simply moving a grain from its position. 

Since each arch is stable on its own right, the arch-based description warrants that any configuration proposed fulfils a basic stability constraint; i.e., that each set of grains identified as an arch has internal contacts that keep it stable. The problem is now moved to generating all possible combination of arches, including how many of them are of a given size, shape, orientation and position and also generating all arrangements of these that can be stable resting on each other. Of course, all of these DoF can be represented with different levels of approximation. 

In Ref. \cite{archensemble}, we have described the five general steps necessary to carry out an Edwards entropy calculation (i.e., the number of states associated to each given volume of the packing) within an arch-based scheme. These are:

\begin{enumerate}
    \item Define the microstate of the system in terms of arches.
    \item Define the external constraints imposed to arches. 
    \item Define a volume function that yields the total volume of the microstate in terms of the arches. 
    \item Define an algorithm to generate all microstates defined in step 1 that comply with the external constraints of step 2. Or sample microstates with equal probabilities. 
    \item Calculate the volume of each microstate generated in step 4 using the function in step 3 and build the DoS.
 \end{enumerate}

Step 4 may constitute a significant limitation to the real possibility of calculating the density of states for systems of reasonable size. Generating all configurations is certainly impractical for most models (especially if they have continuous DoF). This paper demonstrates how to sample configurations by using WL (rather than generating all of them) to accomplish step 4.  

We will focus on a model we have already solved exactly by counting every single possible configurations so as to have a reference system to compare with the WL algorithm results. This is the ``non-interacting rigid arch model'' (NIRA). In this model, only the number of arches $n_i$ of each size $i$ (in number of grains) that form part of the packing is used to describe the microstate. All arches consisting of the same number of grains are considered to occupy the same volume (hence one single possible shape is assumed for each arch size; this is implied in the word ``rigid'' used to name the model). The total volume of the system is assumed to be the sum of the volume of the individual arches and the arch--arch interlocking DoF is not taken into account (hence the word ``non-interacting''). To represent a two-dimensional pack of equal-sized disks, we have taken the volume $v_i$ of an arch of $i$ grains to be the area under the regular polygon that inscribes all disks in a ``regular'' arch. The special 
cases 
of ``arches'' of one particle and two-particle arches are considered separately \cite{archensemble}. The total volume $V[\{n_i\}]$ of the system is

\begin{equation}
\label{volume}
V[\{n_i\}]=\sum_{i=1}^{N}{v_i n_i} \mathrm{\ where\ } v_1=\frac{\sqrt{3} d^2}{2}, \mathrm{\  } v_2=2.1 v_1, 
\end{equation}
\begin{equation*}
v_i= \frac{Nd^2}{4} \tan\frac{\pi}{2N} \left[1+\left(\tan\frac{\pi}{2N}\right)^{-1} \right]^2 \mathrm{\ for\ } i>2.
\end{equation*}
Where $N$ is the total number of disks of diameter $d$ in the system.

It is important to mention that, typically, the maximum size that an arch can take is physically bounded (e.g., due to the size of the container that holds the granular sample). Hence, we will put a cutoff $C$ to the largest arch allowed in the system. The cutoff $C$ is an external constraint (that is imposed in step 2, above). Importantly, this cutoff imposes a limit to the correlations in the system which leads to an extensive entropy (see Ref. \cite{archensemble})

In counting microstates, one has to bear in mind that arches of the same size are indistinguishable in the NIRA model, whereas arches of a different size can be distinguished. Hence, if there are $n_i$ arches of $i$ grains in a configuration (with $i=1,... C$, being $C$ the size cutoff), then the number of permutation of arches that yield distinguishable microstates with these arches is  

\begin{equation}
        N_A!/(n_{1}!...n_{C}!),
    \label{factordeg}
\end{equation}
where $N_A=\sum_i n_i $ is the total number of arches of the configuration (including those ``arches'' of size 1). 

\begin{figure}
\begin{center}
    \includegraphics[width=0.35\textwidth]{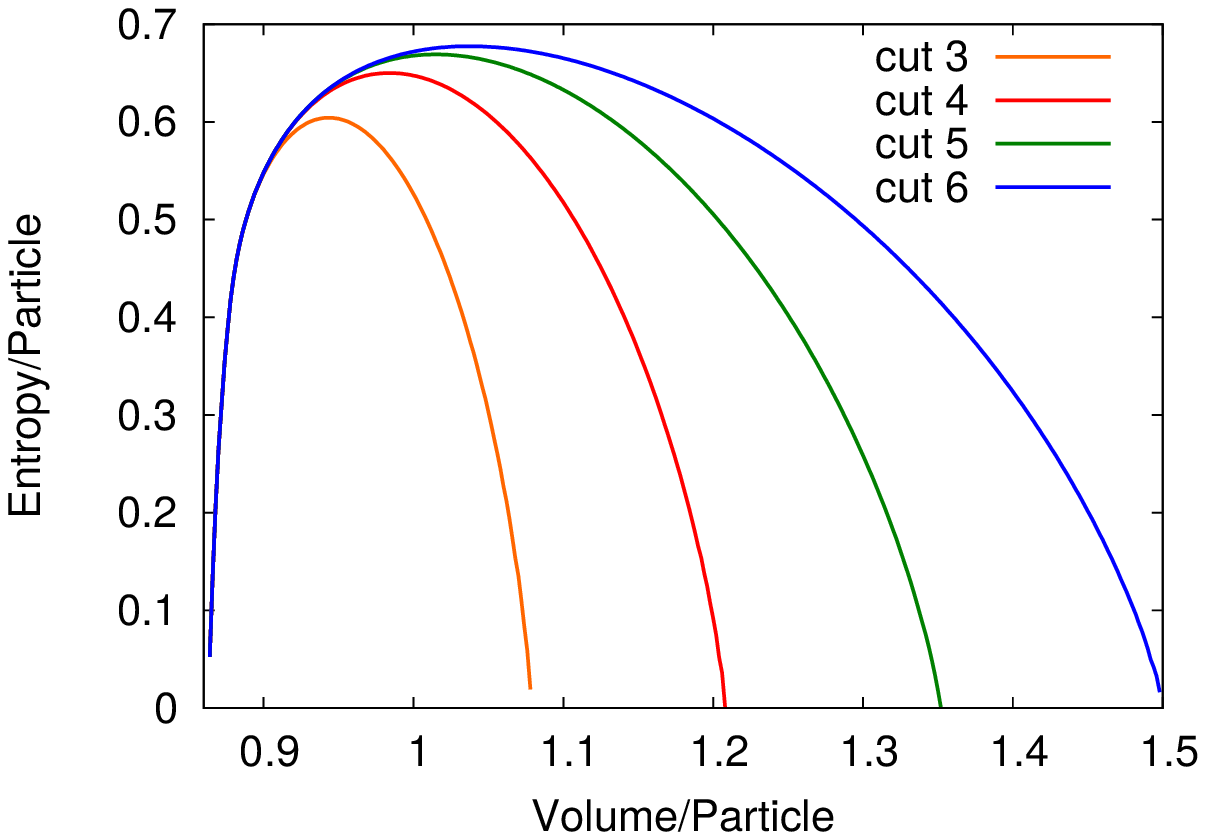} (a)\\
    \includegraphics[width=0.35\textwidth]{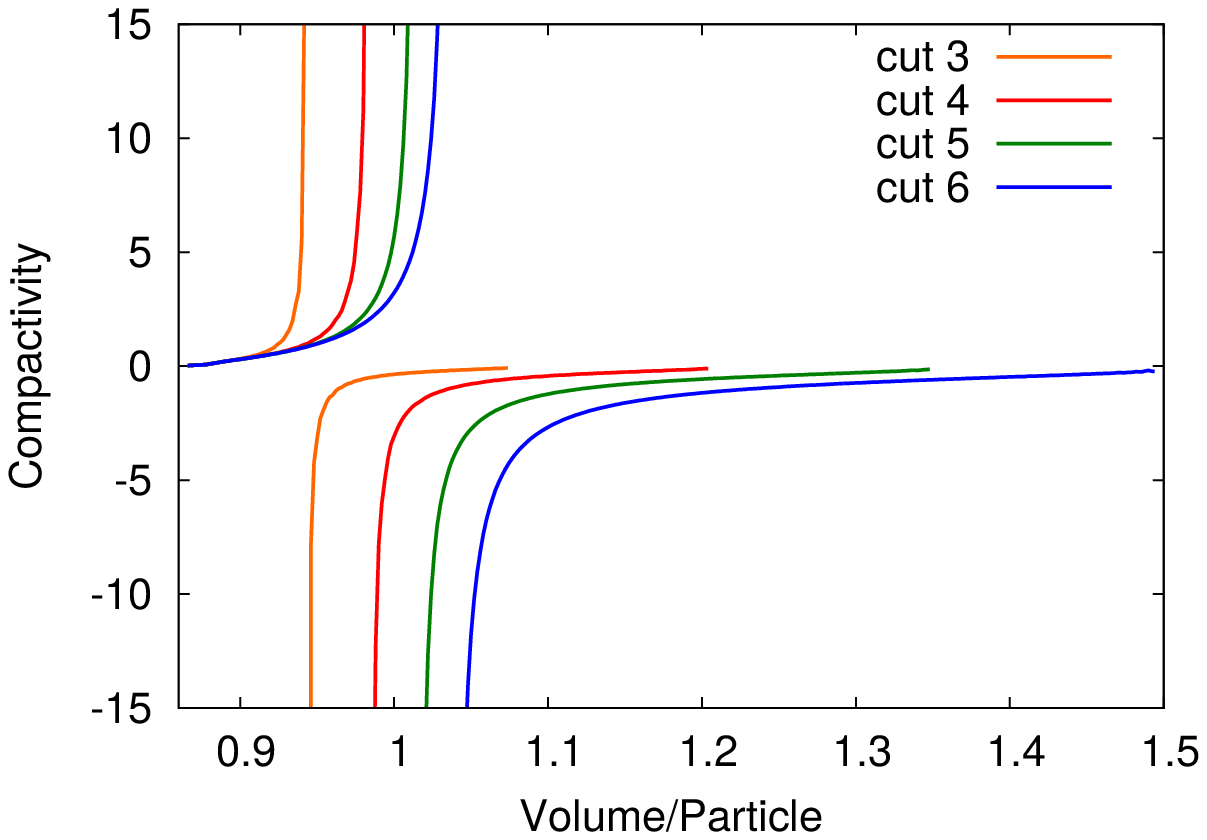} (b) 
\end{center}
\caption{(a) Entropy as a function of volume for the NIRA model in 2D calculated by counting all possible configurations for 500 disks \cite{archensemble} using different arch size cutoffs. (b) The corresponding compactivity calculated by numerical differentiation.}
    \label{entropy}
\end{figure}

Despite all the simplifications, the model applied to a 2D system of equal-sized disks yields qualitative agreement with DEM simulations of tapped disks \cite{archensemble}. The NIRA model is in many 
respects similar to an ideal gas of excitations or quasi-particles (the arches) with a single DoF (their size). 

Figure \ref{entropy}(a) shows the entropy $S$ calculated by counting all possible microstates for $500$ disks using different $C$ for the largest arch allowed \cite{archensemble}. As it is expected, if $C$ increases, looser configurations are possible and hence states with higher volumes become more significant. However, the entropy for low volumes rapidly converges to a well defined curve. Part (b) of Fig. \ref{entropy} shows the compactivity $\chi$ defined as $\chi^{-1}=\partial S / \partial V$. This is the analogue to the temperature in thermal systems \cite{Edwards19891080}. The entropy presents a maximum as observed by others \cite{mcnamara2009measurement,PhysRevLett.101.128001}. States for volumes beyond this maximum correspond to negative compactivities [see Fig. \ref{entropy}(b)]. This is caused by the inversion population of these volume bounded systems. Some authors suggest these negative $\chi$ macrostates may be inaccessible, 
however, this does not need to be the case; some preparation protocols may indeed lead to very low packing fractions \cite{PhysRevLett.101.128001,pugnaloni2008}. An interesting prediction of the NIRA model is that systems constrained by different $C$ achieve the same $\chi$ at different specific volume $V/N$ (i.e., at different packing fractions). As a consequence, two samples of grains ``equilibrated'' to the same compactivity will show distinct packing fractions if the maximum arch size possible in each sample is different. Different values of $C$ in practice may be achieved by using narrow containers or by changing the static friction coefficient of the grains. There have been some progress in the study of the equilibration of vibrated granular samples in ``contact'' \cite{puckett}. However, there are still no attempts to couple static granular packs under gravity. Further developments in this direction may help validating this prediction of the NIRA model.

The configurations of the NIRA model are compatible with different representations. In the following subsections we will discuss some of these representations and their suitability for the implementation of the WL algorithm.

\subsection{The arch size distribution representation}

In our previous paper \cite{archensemble}, we have used a vector $\{n_i\}$ that represents the number of arches consisting of $i$ grains in the configuration, i.e., $\{n_i\}=(n_{1},n_{2},...,n_{C})$, with $C$ the largest arch allowed in the system and $n_1$ the number of grains not forming arches. As an example, a possible configuration in a system of $N=10$ grains and a cutoff arch length of $C=6$ represented in this way could be:
\begin{equation}
        ( 3, 1, 0, 0, 1, 0 ).
    \label{ae}
\end{equation}
In this example, there are three grains not forming arches, two grains forming an arch of size two, and five grains forming another arch of size 5.

In Ref. \cite{archensemble}, we have swept all possible configurations and multiplied each by its analytical degeneration due to the different permutations of arches with repetitions given by Eq. (\ref{factordeg}).

It is difficult to propose an algorithm to move between configurations represented in this way and yet comply with the symmetry of the selection probability required by the WL algorithm. For example, consider a move that consists in removing a grain from one arch of size $k$ and adding it to another arch of size $k'$. Such move would require subtracting 1 from the coordinate $k$ of $\{n_i\}$ and adding 1 to the coordinate $k-1$, since this arch will now be smaller by one grain. Additionally, the coordinate $k'$ of $\{n_i\}$ needs to be reduced in 1 and the coordinate $k'+1$ increased in 1, since this arch will now be part of the set of arches larger by one grain. There are different ways of selecting a grain to be moved and to select its arch of destination. 

Let us consider, for instance, that all grains and destination arches are chosen with same probability. Now consider a move in the Markov chain that takes configuration (\ref{ae}) [i.e., $\mu=(3,1,0,0,1,0)$] into configuration $\nu=(2,1,0,0,0,1)$. This corresponds to taking one grain that was not forming an arch and inserting it in the five-particle arch to make it a six-particle arch. The probability of selecting a particle from an ``arch of size one'' in this case is $3/10$. The probability of choosing the arch of size five as the destination is $1/5$ (there are four other arches in configuration $\mu$ plus the possibility of leaving the grain on its own without forming an arch with others). Hence the selection probability is $S_P(\mu\rightarrow\nu)=3/50$. A similar analysis shows that to return to the original configuration $S_P(\nu\rightarrow\mu)=6/10\times 1/4=3/20$ (there are 6 grains out of ten that can be taken from the six-grain arch and there are three possible other 
destination arches plus the case 
with the grain not forming an arch in the new configuration). Clearly, this selection probabilities are non-symmetric as they should be to apply the algorithm of section II.

A possible workaround to the previous representation is to multiply each coordinate of the vector $\{n_i\}$ by the corresponding arch size. Therefore, each coordinate now indicates how many grains are involved in all arches of the given size. In this new representation, the configuration of Eq. (\ref{ae}) (10 particles with a cutoff $C=6$) is written as
\begin{equation}
       \{n'_i\} = (3,2,0,0,5,0)
    \label{ae1}
\end{equation}

In this case, a trial move may consist in randomly transferring a grain from one coordinate $k$ to another coordinate $k'$. This is done simply by subtracting 1 to $\{n'_k\}$ and adding 1 to $\{n'_{k'}\}$. In this representation, the selection probability of moving a particle from one arch of size $k$ to another of size $k'$ is $S_P(\mu\rightarrow\nu)= 1/N \times 1/(C-1)$ irrespective of $k$ and $k'$. Hence, the selection probability is symmetric and suitable to implement the entropic sampling using WL. 

Unfortunately, the representation of Eq. (\ref{ae1}) does not tell apart two microstates that differ only in the permutation of two arches of different sizes. Therefore, the corresponding degeneracy given by Eq. (\ref{factordeg}) cannot be accounted for \footnote{One is tempted to add to the degeneracy factor (\ref{factordeg}) to correct the entropy in step III of the algorithm. However, the algorithm compensates this factor in order to obtain a flat histogram. Therefore this is not a viable solution.}. More importantly, the change of representations from Eq. (\ref{ae}) to Eq. (\ref{ae1}) is not an exact mapping because these newly proposed moves allow for ``fractions of arches'' to exist since we do not request that the coordinates of $\{n'_i\}$ be a multiple of $i$ after each move. For instance, the configuration $(0,0,0,0,0,10)$ is allowed in representation (\ref{ae1})  but does not exist in representation (\ref{ae}). In the new representation such configuration implies that there is one arch of size $6$ 
plus a fraction of an arch of size six. We can still treat these ``fracrtions of arches'' by assigning to them a fraction of the arch volume. However, there is a large number of new unrealistic configurations (there are many ways of chosing a numbers that are not conmensurate with the arch size) in this representation that will bias the result for the entropy. 

\begin{figure}
    \includegraphics[width=0.45\textwidth]{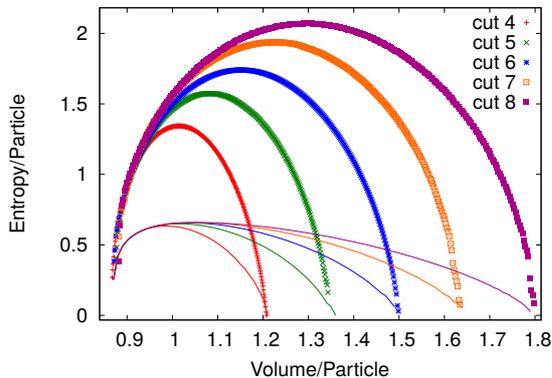}
    \caption{Entropy as a function of volume for the NIRA model calculated using the WL algorithm (symbols) for representation (\ref{ae1}) for $4<C<8$. The full lines represent the exact results for $200$ grains.}
    \label{nodegeneration}
\end{figure}

Figure \ref{nodegeneration} shows the entropy for the NIRA model using representation (\ref{ae1}) for different maximum arch sizes $C$ compared with the exact result obtained by counting all configurations and all permutations \cite{archensemble}. As we can see, not including all distinguishable permutations and including the new ``fractional arches'' gives a wrong entropy function. 

\subsection{The arch listing representation}

As previously discussed, other ways of representing the system should be carefully chosen in order to ensure that there exist moves that sample the different configurations uniformly. 
An alternative natural representation consists of using a vector, $\{m_i\}$, with $N$ coordinates, where each coordinate $m_i$ can take any value from $0$ to $C$, the cutoff for the arch size, provided that $\sum_i m_i = N$. The content of each coordinate indicates that the configuration has an arch of that size. The configuration of Eq. (\ref{ae}) in these representations can be expressed, for example, as

\begin{equation}
    ( 5,0,0,1,0,1,2,0,1,0 ). \label{arch-list}
\end{equation}
There are, of course, multiple permutations in Eq. (\ref{arch-list}) that lead to the same distribution of arch sizes compatible with Eq. (\ref{ae}).

Trial moves in the system represented in this way can be done by subtracting one from a non-zero $m_i$ and adding one to any other coordinate that has a value smaller than $C$. This is equivalent to reducing the size of one arch in one particle and either creating a new arch of size one (if the new coordinate had a zero value) or increasing the size of another arch. Unfortunately, this algorithm has a non-symmetric selection probability. $S_P= 1/N_A \times 1/(N-N_C)$, where $N_A$ is the number of arches (i.e., the number of non-zero coordinates in the configuration) and $N_C$ is the number of arches with the maximum allowed size $C$. Therefore, $S_P$ will depend on the total number of arches and the number of arches of size $C$.

Besides the non-symmetric $S_P$, the system represented in this way and sampled with these trial moves clearly overestimates the number of states of a given volume. This is because, apart from the $N_A!/(n_1!...n_C!)$ permutations of the distinguishable arches [see Eq. (\ref{factordeg})], there are $N!/(N-N_A)!$ additional permutations due to the zeros in a given vector in Eq. (\ref{arch-list}) (there are $N-N_A$ zeros). 

One can, in principle, resolve these issues. The selection probability may be turned into a symmetric one by adapting the acceptance probability in Eq. (\ref{aceptance}). The degeneracy due to the presence of zeros in the representation (\ref{arch-list}) can be handled by switching to a representation without including these, and allowing for a vector of variable length. However, although less intuitive, there is a much simpler, suitable representation that we discuss in the next section.

\subsection{The binary arch representation}

\begin{figure}
    \includegraphics[width=0.45\textwidth]{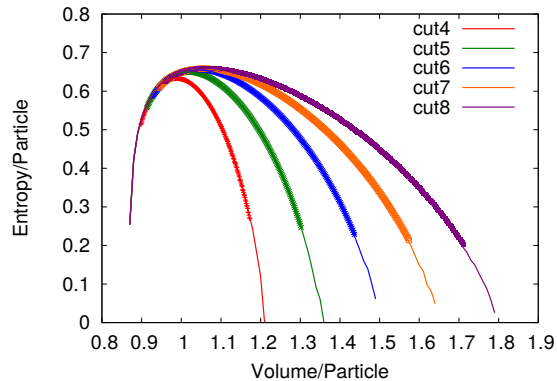}
    \caption{Entropy as a function of volume for the NIRA model with cutoff $4 \leq C \leq 8$ (symbols as in Fig. \ref{nodegeneration}) using the binary arch representation to carry out the entropic sampling through the WL algorithm for $200$ grains. The solid lines correspond to the exact counting of microstates from Ref. \cite{archensemble}.}
    \label{exact}
\end{figure}

Finally, we present a representation that comply with the symmetric selection probability and simultaneously account for the permutation of distinguishable arches.

In this case, the system is chosen to be represented by a vector of $N$ coordinates with binary values (zeros and ones). These coordinates are not associated to specific particles 1, 2, 3, etc. Rather, as we move along the vector from left to right, we can think the ones as representing a first grain in an arch (whatever its identity) and the following zeroes as the remaining particles. For instance, the configuration in equation \ref{ae} in this new representation can be given by
\begin{equation}\label{final}
    (\underbrace{1,0,0,0,0}_{5},\underbrace{1}_{1},\underbrace{1}_{1},\underbrace{1,0}_{2},\underbrace{1}_{1}).
\end{equation}
In this representation, all permutations of arches of different sizes are accounted for naturally. The sections of the vector representing an arch [underbraced in Eq. (\ref{final})] can be permuted to yield all distinguishable configurations [see Eq. (\ref{factordeg})], which correspond to different vectors in this binary representation. Indistinguishable configurations corresponding to permutations of arches of same size are also indistinguishable for the binary vector.

The number $N_A$ of arches in a configuration is simply the sum of all the elements of the vector. Note that the underbraced numbers coincide in value and order with the non-zero figures of the vector described by Eq. (\ref{arch-list}).

The trial moves consist in picking a coordinate and changing the state of that coordinate (if 1 change to 0 and \textit{vice versa}). This results in a symmetric selection probability of $S_P(\mu\rightarrow\nu) = S_P(\nu\rightarrow\mu) = 1/N$. In each move, the constraint imposed by the cutoff $C$ must be checked and the trial configuration must be rejected whenever the constraint is not complied with.

In Fig. \ref{exact}, the result for this representation and sampling strategy is plotted along with the exact result showing a remarkable agreement. 
%

%

\section{Conclusions}
We have been able to compute the entropy of a system of non-interacting rigid arches using a WL algorithm in the volume ensemble in different representations.

We have exposed the difficulties in dealing with different representations of the configurations of arches and the mechanisms used to propose trial moves for the WL algorithm. These difficulties appear during the choice of a simple sampling scheme that ensure a symmetric selection probability of configurations. Additionally, the degeneracy due to distinguishable permutations of arches pose a further complication in the use of WL. The most suitable representation that we found for a non-interacting system of rigid arches resulted in a binary vector.

We believe that entropic sampling of arches through the WL algorithm has a great potential for testing the granular statistical mechanics hypothesis (such as equiprobability and ergodicity). Having a sampling algorithm like WL adapted for these types of models is crucial to continue the road map towards the refinement of an arch-based framework for static granular packs. In particular, the non-interacting condition is clearly a crude approximation and should be lifted, along with the introduction of a more accurate volume function.

\end{document}